# Structural, vibrational and transport properties of compound forming liquid Li-Bi alloys

S.G. Khambholja[1]*, A. Abhishek[2] and B.Y. Thakore[3]

[1]Department of Science & Humanities, B & B Institute of Technology, Vallabh Vidyanagar, Gujarat, India

[2]Institute of Plasma Research, Gandhinagar, Gujarat, India

[3]Department of Physics, Sardar Patel University, Vallabh Vidyanagar, Gujarat, India

*Corresponding author: *physik.shyam@gmail.com*

**Abstract**

Due to the compound forming tendency, some of the liquid metal alloys show anomalous behavior in their physical and chemical properties. Near the compound forming concentration, their electrical resistivity is beyond the metallic values and hence they may be labelled as liquid semiconductors. Lithium-Bismuth is one such system. It shows some interesting features in terms of physical and chemical properties such as departure from nearly free electron theory, very high value of electrical resistivity (~ 2000 $\mu\Omega$ cm) near the compound forming composition. While dealing with the electrical resistivity of liquid alloys with very high values of electrical resistivity, the famously used approaches such as Faber-Ziman theory and Morgan theory have some limitations. Hence, some modifications in these theoretical formalisms are required in order to reproduce the experimental values of the electrical transport properties. We, in the present work have modeled liquid Li-Bi system using model potential formalism in conjunction with the established theoretical models along with suitable modifications to study structural, elastic and transport properties. In particular, we have treated the effective valence of pure Li and Bi as parameters and we have calculated the phase shifts using model potentials rather than muffin-tin potential. The results are compared with the results of molecular dynamics simulation and other theoretical models. It is observed that the t-matrix formulation in conjunction with the model potential formalism is able to reproduce the correct trends in the electrical resistivity isotherm. Whereas the results of Faber-Ziman and Morgan theory are highly underestimated, the non-metallic behavior near the *critical composition* can be explained clearly from the present results of electrical resistivity. Further, phonon frequencies and sound velocities are also estimated.

## 1. Introduction

Lead-Bismuth eutectic (LBE) is one of the important materials being considered for application in fast neutron reactors like Russian SVBR 100, Belgian MYRRHA etc as first circuit coolant due to their exceptional properties such as low melting point, high boiling point, excellent thermal conductivity, chemical inertness, low neutron absorption and comparatively better performance even in irradiated situations [1]. On the other hand, the known limitation is that LBE is corrosion aggressive towards iron-based materials [1]. In this situation, it will be really interesting to study structural properties of LBE. Li-Bi based Pb-Li-Bi alloy are also one novel material for its application as a coolant in fusion reactors [2]. These alloys are known for their marked chemical order and heterocoordination tendency due to strong difference between valences of pure Li and Bi [3]. The study of electrical transport properties of Li-Bi alloys indicates strong departure from simple metallic behavior [3-5]. Moreover, the Li-Bi system contains two elements Li and Bi with disparate masses, the mass ratio being ~30.10. This high mass ratio is considered responsible for the peculiar behavior of collective excitations [3, 6]. The system is somehow similar in terms of its features to liquid lead-lithium (Pb-Li) alloys. Recently, we have carried out the study of structural, electric transport and vibrational properties of liquid Pb-Li alloys in a wide range of temperature and Li concentration [7-12]. From the study of electrical resistivity of liquid Pb-Li alloys, we have concluded that we need to go beyond the popularly used Faber-Ziman (FZ) approach as well as the Morgan theory to compute the electrical resistivity of compound forming Pb-Li alloys. Wax et al [3] have carried out multiscale study of various physical properties of Li-Bi alloys in three different Bi concentrations ($x_{Bi}$ = 0.70, 0.43, 0.30) using empirical oscillating pair potentials (EOPP). From their study, it is clearly seen that the first minimum in the pair potential in case of $Li_{30}Bi_{70}$ alloy is positive and it is negative in the case of $Li_{57}Bi_{43}$ and $Li_{70}Bi_{30}$ alloys. It is also observed by Mayou et al [13] that $x_{Bi}$ = 0.25 is the *critical* composition for Li-Bi alloys. Near this composition, anomalies in the electrical conductivity and volume contraction are observed. In their study of phase diagram, Dorini et al [14] have noted that molar free energy of mixing and molar enthalpy of mixing show minima near eutectic composition. Further, they have reported that $Li_3Bi$ is a strongly stable compound. Wax et al [3] have noted that a pseudogap appears at the Fermi level in the electronic density of states of Li-Bi alloys and it deepens as Bi concentration decreases. They have attributed the high value of electrical resistivity (~ 2000 μΩ cm) of $Li_3Bi$ liquids to the presence of the pseduogap. Thus, it is clear that it will be interesting to study electrical resistivity of liquid Li-Bi alloys particularly near the *critical* composition using theoretical approach, which will not only provide us the validity of the model used but can be helpful to understand the charge transfer mechanism. Since, the structure factors are important ingredients to study electrical transport properties of alloys, and since such a study may also be helpful to understand concentration fluctuations in the long wavelength limit, we have computed the structure factors too. The study of interatomic pair potential is further helpful to understand the nature of ion-ion interaction. On the other hand, the study of phonon frequencies, sound velocity and electrical transport properties such as electrical resistivity is found less in literature. Motivated by this scenario, we in the present work have carried out the study of structural, vibrational and transport properties of liquid Li-Bi alloys.

## 2. Method of Computation

### 2.1 Pair potential, phonon spectra and sound velocity

Wax et al [3] have carried out multiscale modelling of liquid Li-Bi alloys and have observed that the second order perturbation theory may give poor description of properties for compound forming alloys showing departure from simple nature. However, we in the present work have calculated effective pair potential using second order perturbation theory [15] due to many reasons. First of all, this approach allows us to use various forms of electron-ion potentials as well as local field correction functions. Secondly, it allows us to fit the computed pair potential with results of molecular dynamics (MD) simulation. Further, since we have extended our present work to calculate electrical resistivity of liquid Li-Bi alloys using same model potential coupled with the t-matrix formulation, the role of model potential becomes important and its validity has to be determined. Wax et al [3] have used EOPP along with classical molecular dynamics simulation to simulate various properties of Li-Bi alloys using experimental values of alloy density. Further, they have obtained the results by fitting the relevant properties with the results of Vienna Ab initio Simulation Package (VASP) simulation. While using second order perturbation theory, the choice of electron-ion potential and screening function is crucial in determining material properties at a given temperature. In the present work, we have used empty core model (ECM) due to Ashcroft [16] along with local field correction function due to Ichimaru and Utsumi [17] to calculate the inter-ionic potential. The single parametric ECM potential is simple in nature. The only parameter of the potential namely the core radius is calculated from the known density or it can be obtained from the fitting procedure. The value of the core radius in the present work is 0.5768 Å.

The inter-ionic potential using the second order perturbation theory in the Wills-Harrison (WH) form can be written as following [15].

$$V(r) = \frac{Z^2 e^2}{r} + \frac{\Omega_0}{\pi^2} \int F(q) \frac{\sin(qr)}{qr} q^2 dq \tag{1}$$

Here, the first term on the right side arises due to the direct coulombic interaction between the ions and it is purely repulsive. The second term indicates indirect interaction between ions via electron cloud and it is attractive one. $F(q)$ is the energy wave number characteristics. In the present work, we have estimated energy wave number characteristics using model potential formalism. Further, we have obtained the core radius of the alloy by fitting the presently calculated effective pair potential with the results of MD simulation of Wax et al [3]. Thus, although we have used second order perturbation theory, a fitting procedure allows us to obtain a form of pair potential which is in agreement with the results of MD simulation. Having obtain effective pair potential, the characteristic frequency or Einstein frequency of the system can be obtained using phenomenological approach of Hubbard and Beeby (HB) [18] as following.

$$\omega_E^2 = \frac{4\pi\rho}{3M} \int_0^\infty g(r) V''(r) r^2 dr \tag{2}$$

Here, g(r) is the radial distribution function (RDF). In the present work, we have used the values of RDF reported by Wax et al [3]. Einstein frequency can be further used to calculate longitudinal and transverse phonon frequencies using following set of equations [18].

$$\omega_L^2(q) = \omega_E^2 \left[ 1 - \frac{\sin(q\sigma)}{q\sigma} - \frac{6\cos(q\sigma)}{(q\sigma)^2} + \frac{6\sin(q\sigma)}{(q\sigma)^3} \right] \tag{3}$$

$$\omega_T^2(q) = \omega_E^2 \left[ 1 - \frac{3\cos(q\sigma)}{(q\sigma)^2} + \frac{3\sin(q\sigma)}{(q\sigma)^3} \right] \tag{4}$$

Finally, sound velocities are calculated from the long wavelength limit of eq. (3) and (4).

**2. Structure factor and electrical resistivity**

Akinlade et al [19] have noted that in order to explain the electrical resistivity of Li based alloys with high valence of the solute, one has to go beyond the Faber-Ziman (FZ) model

[19] to explain the high values of electrical resistivity. Wax et al [3] have reported that the Li rich Li-Bi alloys have compound formation tendency and are far from nearly free electron model. Thus, it is to be believed that the most common approaches used to compute electrical resistivity of liquid alloys namely Faber-Ziman (FZ) approach [20] cannot reproduce the experimental values in the resistivity isotherm. In most of the cases, the values of electrical resistivity calculated using FZ approach are smaller compared to other theoretical approaches namely $2k_F$ scattering model [21] and t-matrix formulation [22] as well as experimental results. Whereas the $2k_F$ scattering model [21] requires the calculation of finite mean free path, it has limited validity when the interatomic spacing is more than the mean free path. On the other hand, the t-matrix formulation [22] requires the calculation of phase shifts using muffin-tin (MT) potential [23]. The experimentally observed value of electrical resistivity of Li-Bi near the *critical* composition is around 2000 μΩ cm [3, 5] and thus it is clear that none of the above-mentioned approach can reach up to this much high value of electrical resistivity as both the approaches are relatively simple. Most recently, we have developed a method of calculating electrical resistivity of liquid Pb-Li alloys using model potential formalism coupled with the t-matrix formulation [22]. In this approach, following modifications are proposed.

1. We have calculated the finite phase shifts of pure Li and Bi up to second order using model potential formalism instead of (MT) potential [23], which in turn are used to calculate electrical resistivity. As discussed above, we have use ECM in the present work.
2. We have treated the effective valence of pure Li and Bi as parameters to obtain the experimental values of electrical resistivity of pure Li and Bi at the desired temperature.
3. We have assumed that the calculated phase shifts and t-matrix form factors of pure Li and Bi are independent of alloy concentration.
4. We have neglected the contribution from band energy.

The second modification is similar to the approach Esposito *et al* [24], who have suggested to determine the fermi energy in a self-consistent way using the concept of "*effective conduction electrons*". Through the analysis of band structure of rare earth metals, Delley et al [25] suggested that *effective conduction electrons are* different from valence of the element. Further, the reason behind treating the effective valence as a parameter is associated with the observation of anomaly in the density of states near fermi surface, which in turn reduces the effective valence. Thus, in the present work, we have calculated the electrical resistivity of

Li-Bi alloy using the concept of *effective conduction electrons* or *effective valence* of pure Li and Bi rather than the actual valence. More theoretical details and justification about present approach can be found in our previous work [8]. Since, Pb-Li and Li-Bi alloy have similar properties viz. departure from free electron model, high mass ratio and almost similar valence ratio, we have extended our method of calculating electrical resistivity of Pb-Li alloys to Li-Bi alloys.

In the t-matrix formulation, electrical resistivity of liquid binary alloys is calculated using following equation [22].

$$\rho_{el} = \frac{\pi \cdot \Omega_0}{k_F^2} \int_0^1 \left(\frac{q}{2k_F}\right)^3 d\left(\frac{q}{2k_F}\right) |T_{alloy}|^2 \quad (5)$$

Here, $|T_{alloy}|^2$ is the total t-matrix form factor and is given as following.

$$|T_{alloy}|^2 = c_i \cdot |t_i|^2 [1 - c_i + c_i a_{ii}(q)] + c_j \cdot |t_j|^2 [1 - c_j + c_j a_{jj}(q)] + c_i c_j (t_i^* t_j + t_i t_j^*)[a_{ij}(q) - 1] \quad (6)$$

Here, $t_i$ and $t_j$ are the t-matrix form factors of pure element forming alloys [in the present case Li and Bi]. $a_{ii}(q)$, $a_{jj}(q)$ and $a_{ij}(q)$ are the Faber-Ziman partial structure factors [26] corresponding to Li-Li, Bi-Bi and Li-Bi, respectively. $c_i$ is the concentration of Li and $c_j$ is the concentration of Bi such that $c_i + c_j = 1$. t-matrix form factors of pure elements are calculated using following equation.

$$t(E_F, q) = -\frac{2\pi}{\Omega_0 (2E_F)^{1/2}} \sum_l (2l+1) \sin \eta_l \exp(i \cdot \eta_l) P_l(\cos\theta) \quad (7)$$

In determining phase shifts, we have used Ashcroft empty core model (ECM) potential [16] to incorporate ion-electron interaction in conjunction with Percus-Yevick (PY) hard sphere structure factors [26] and experimental values of densities of both Li [27] and Bi [28]. Once the t-matrix form factors of pure Li and Bi are obtained, they are used along with Faber-Ziman partial structure factors [26] in equation (6) and finally equation (5) is used to compute electrical resistivity as a function of Bi concentration. All the calculations are carried out at 1073K.

## 3. Results and Discussion

Presently calculated effective pair potential of $Li_{70}Bi_{30}$ alloy is shown in Fig. 1 along with results of Wax et al [3]. It is seen from the Fig. 1 that the presently calculated results are in good agreement with the results of MD simulation. In particular, the oscillating nature of the pair potential are in phase with each other near the first minimum. EOPP results of Wax et al

[3] exhibit oscillating nature in the entire range. On the other hand, the presently calculated pair potential becomes relatively flat beyond first neighbor distance. Moreover, in the short r region, presently calculated pair potential is more repulsive in nature as compared to the EOPP results and in the high r region, the scenario is reversed. The minimum in the presently calculated effective pair potential is negative around the first neighbor distance, which is in agreement with the EOPP results. Presently calculated values of longitudinal and transverse phonon frequencies are plotted in Fig. 2. The calculated values of phonon frequencies display all the broad features of collective excitations in disordered systems. Presently calculated longitudinal phonon frequency shows first minimum at about 1.43 Å$^{-1}$. On the other hand, the first peak in the experimental value of total structure factor reported in [3] is observed at around 1.48 Å$^{-1}$. Both the values are close to each other, as expected. Presently calculated Einstein frequency comes out to be 12.57 THz, for which no data is available for comparison in literature. Further, the presently calculated longitudinal and transverse sound velocity turn out to be $1.89 \times 10^5$ m/s and $1.09 \times 10^5$ m/s.

Presently computed t-matrix form factor for pure Li and Bi are shown in Fig. 3. Further, presently calculated FZ total structure factor is shown in Fig. 4 along with the experimental results. One of the reasons behind error in the theoretical calculations of electrical resistivity is the error in the determination of structure factor. In the present work, we have calculated the FZ structure factor from the Ashcroft-Langreth structure factors with a total packing faction 0.45. Since, Li-Bi is not purely additive in the entire concentration range, a difference between the presently calculated total structure factor shows some deviation from the experimental value.

Presently calculated energy dependent resistivity of pure Li and Bi are shown in Fig.5 along with the results of Khalouk et al [29]. The authors in [29] have determined the energy dependent resistivity of pure Bi at 1223K. On the other hand, present calculation is carried out at 1073K. Thus, a difference in the results in Fig. 5 is obvious. The resistivity initially increases and beyond the $q = 2k_F$, it starts decreasing. Fig. 5 also shows that the electrical resistivity is strongly dependent on the choice of $k_F$.

Fig. 6 shows the presently calculated values of electrical resistivity as a function of Bi concentration along with the results of FZ theory and Morgan theory as reported in [19], experimental value near critical composition and the results of Mayou et al [13]. From, the experimental value, it is clear that Li-Bi alloy shows non-metallic behavior near the critical composition. Further, it is clear that the results of FZ theory and Morgan theory reported by Akilande et al [19] are largely underestimated compared to the experimental value ~2000

µΩ.cm. The experimental value is available only at the critical composition i.e. $Li_3Bi$. The presently calculated value of electrical resistivity near the critical composition is very close to the experimental value. From the presently calculated electrical resistivity, it is clear that near the critical composition, electrical resistivity has a maximum value and it decreases as Bi concentration increases.

## 4. Conclusion

In the present work, we have proposed a method of determining effective valence of pure elements to calculate the electrical resistivity of compound forming Li-Bi alloys by combining model potential formalism with the t-matrix formulation. Due to a pseudogap in the electronic structure, Li-Bi alloy has non-metallic properties near the critical composition. Whereas the results of FZ theory and Morgan theory are underestimated compared to the experimental values, presently calculated results are in far better agreement with the experimental results, particularly near to critical composition. Present results also clearly indicate an asymmetry in the electrical resistivity isotherm. Since, the resistivity calculated using other theoretical approaches namely FZ model and Morgan theory have limitations while dealing with the compound forming alloys like Li-Bi and since the present results not only reproduce the experimental value but also reproduce the trend correctly, we believe that the present approach i.e. coupling model potential formalism with t-matrix formulation is useful to calculate the electrical resistivity of compound forming alloys. The calculated effective pair potential is also in good agreement with the results of MD simulation. The phonon frequencies and sound velocities are estimated for the first time in this work.

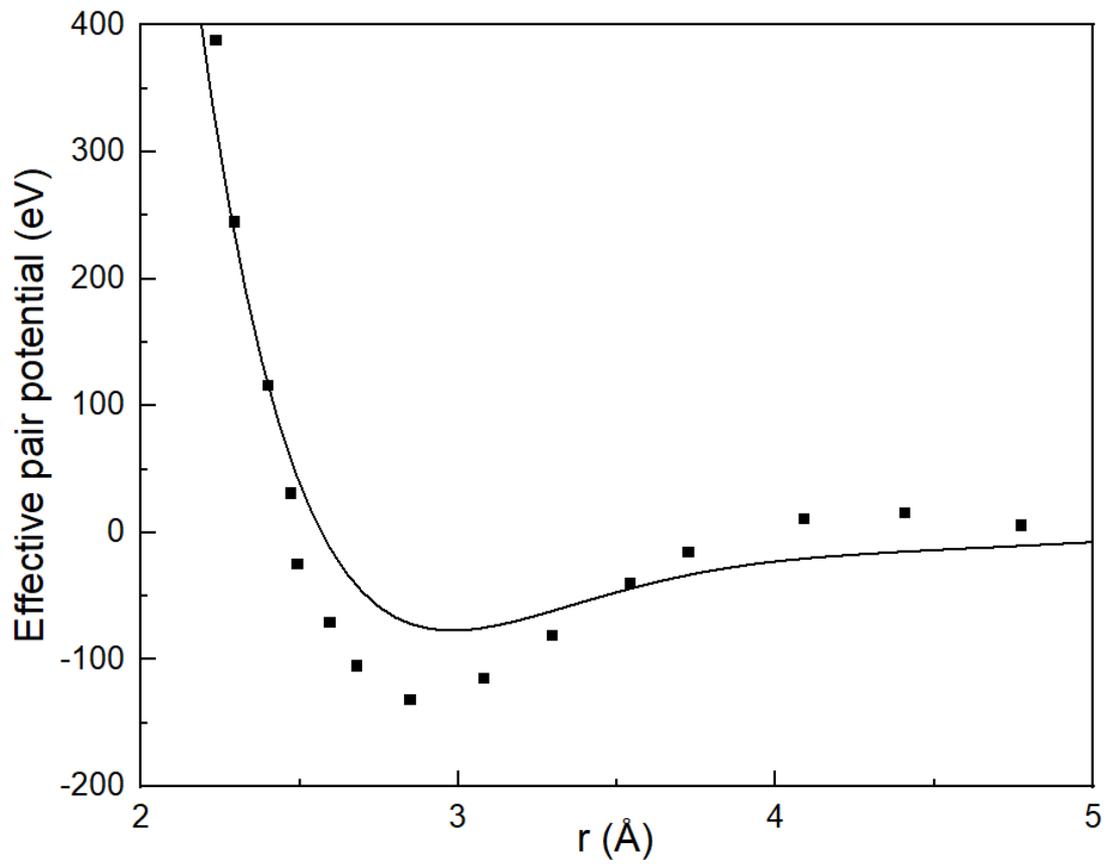

Figure 1. Presently calculated effective pair potential (full line) along with the MD results of Wax et al (symbols) [3].

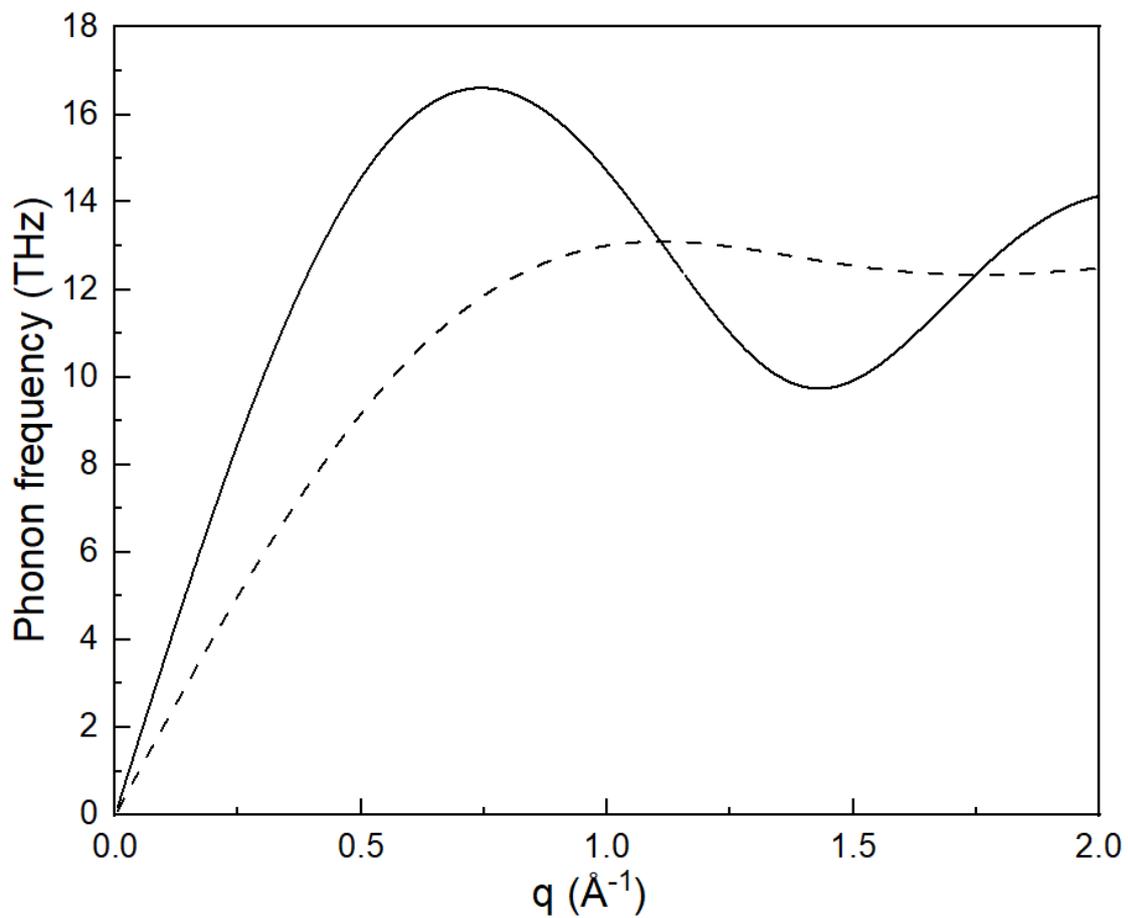

Figure 2. Presently calculated longitudinal (full line) and transverse (dashed line) phonon frequencies.

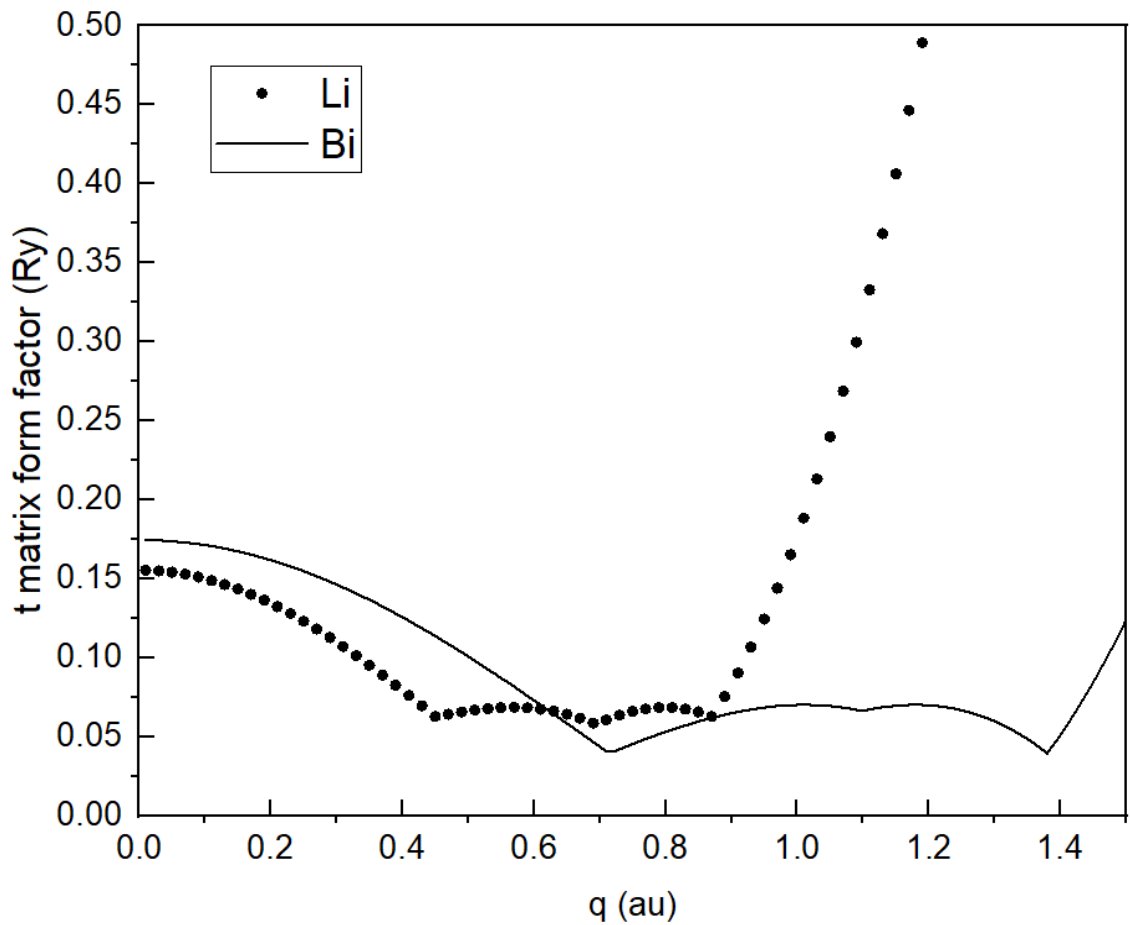

Figure 3. Presently calculated t-matrix form factors of pure liquid Li (dots) and Bi (full line).

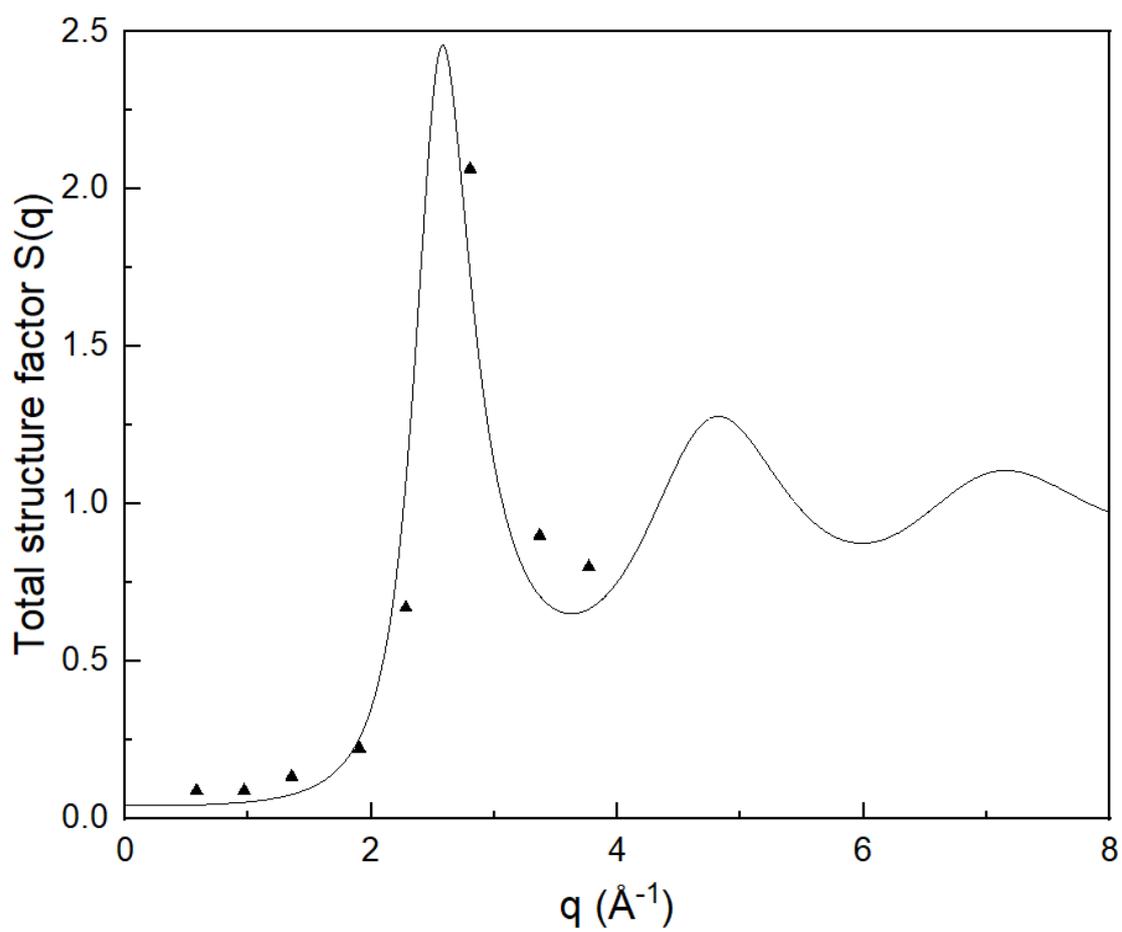

Figure 4. Presently calculated Faber-Ziman total structure factor along with the experimental results (symbols) [3].

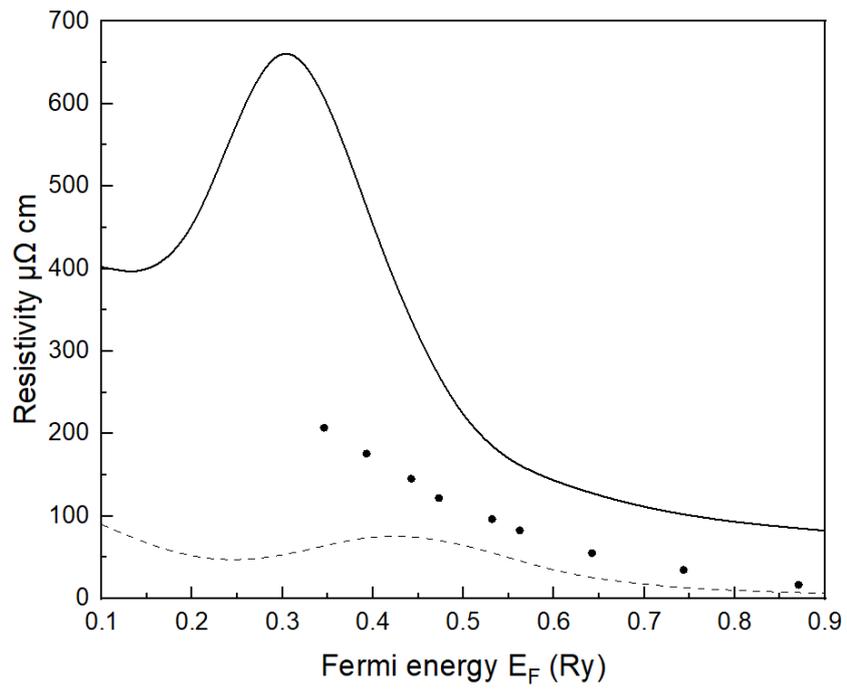

Figure 5. Presently calculated energy dependent electrical resistivity of liquid Li (dashed line) and Bi (full line) at 1073K temperature along with the experimental results for Bi (symbols) at 1223K [29].

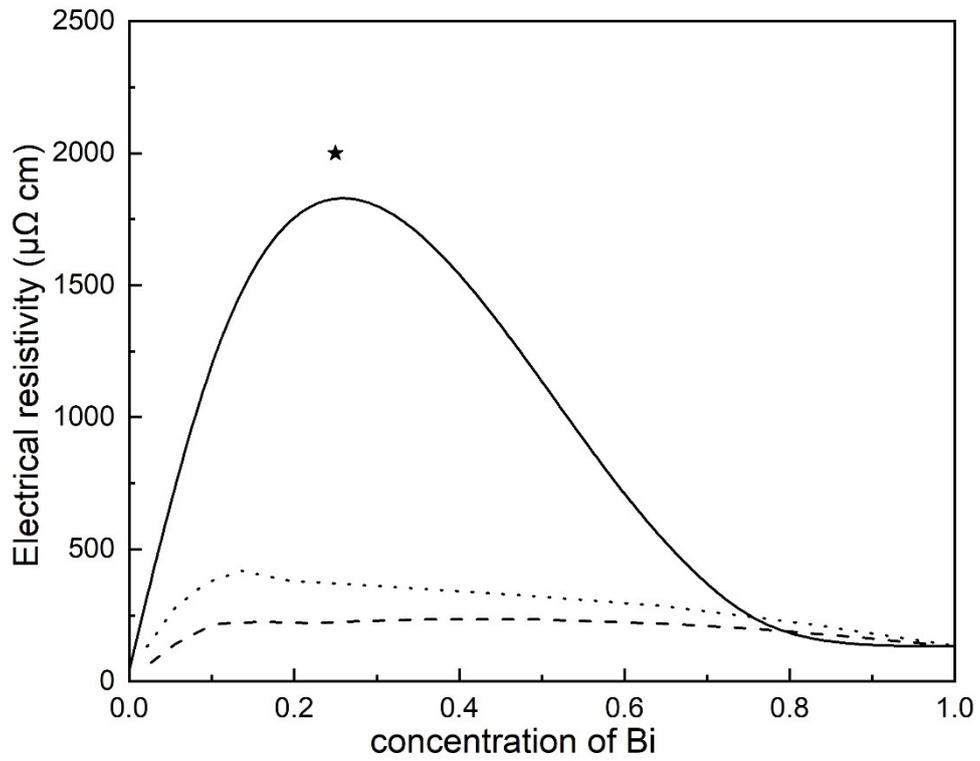

Figure 6. Presently computed electrical resistivity (full line) along with results of FZ model (dashed line) and Morgan theory (dotted line) [19] and a converted experimental value (star) from [3].